\documentclass[aps, prl, twocolumn, showpacs, floatfix,10pt,superscriptaddress]{revtex4-2}

\usepackage[T1]{fontenc}
\usepackage{amsmath}
\usepackage{amsfonts}
\usepackage{amsbsy}
\usepackage{amssymb}
\usepackage{graphicx}
\usepackage{textcomp}
\usepackage[caption=false,singlelinecheck=false]{subfig}
\usepackage{xcolor}
\usepackage{mathrsfs}
\usepackage{mathtools}
\usepackage{bm}
\usepackage{gensymb}
\usepackage{braket,wasysym}
\usepackage[colorlinks,linkcolor=blue,citecolor=red,filecolor=magenta,urlcolor=red,breaklinks]{hyperref}
\usepackage[bottom]{footmisc}
\usepackage{verbatim}
\hypersetup{colorlinks=true, urlcolor=blue, citecolor=red, pdfborder={0 0 0}}
\usepackage{breakurl}
\usepackage{natbib}

\allowdisplaybreaks

\newcommand{\tcm}[1]{\textcolor{magenta}{#1}}

\begin{document}
\title{All-to-All interactions via multifractal wavefunction geometry}

\author{YouYoung Joung}
\email{young_sl.physics@kaist.ac.kr}
\affiliation{Department of Physics, Korea Advanced Institute of Science and Technology, Daejeon, 34141, Korea}
\author{Jemin Park}
\email{physjmp@kaist.ac.kr}
\affiliation{Department of Physics, Korea Advanced Institute of Science and Technology, Daejeon, 34141, Korea}
\author{SungBin Lee}
\email{sungbin@kaist.ac.kr}
\affiliation{Department of Physics, Korea Advanced Institute of Science and Technology, Daejeon, 34141, Korea}

\begin{abstract}
We uncover a generic mechanism through which the intrinsic geometry of multifractal quantum wavefunctions generates effective all-to-all interactions in many-body systems. 
By analyzing the multifractal spectrum, we demonstrate that the simultaneous participation of widely separated length scales creates a global connectivity that bypasses local interaction constraints. 
This nonlocality leads to fast information scrambling, evidenced by sharp changes in the quenched dynamics of the quantum Fisher information and bipartite mutual information with the onset of negative tripartite mutual information.
Such rapid scrambling is a defining feature of strongly chaotic quantum dynamics, and our results identify the systems with multifractal states as a promising solid-state platform for realizing this regime. More broadly, they reveal a new paradigm in which complex, multiscale wavefunction structure intrinsically generates long-range connectivity, providing a natural route to achieving nonlocal behavior in strongly correlated quantum materials.
\end{abstract}

\maketitle

\textit{\tcm{Introduction---}}
Quantum systems with all-to-all interactions exhibit exotic non-equilibrium dynamics that underlie maximal quantum chaos \cite{chen2019quantum, maldacena2016bound, herpich2020stochastic} and enable non-local quantum error correction \cite{choi2020quantum, kong2022near}. Such physics lies at the heart of modern theoretical frameworks, most notably the Sachdev-Ye-Kitaev (SYK) model \cite{maldacena2016remarks, PhysRevLett.70.3339, sachdev2015bekenstein}. Yet in conventional condensed-matter settings, interactions obey locality, with coupling strengths that fall off with distance, making the realization of genuine all-to-all connectivity extremely challenging \cite{
campa2014physics, defenu2023long}.  Although engineered platforms such as random circuits or disordered quantum dots can partially disrupt locality \cite{
lechner2015quantum, chen2018quantum, chew2017approximating}, achieving full connectivity arising solely from a material’s intrinsic properties remains an open challenge. We suggest that systems possessing complex, multiscale structures are the natural candidates to overcome this fundamental constraint.

The multi-scale paradigm finds its prime example in multifractality. Defined by a continuous spectrum of scaling exponents, this phenomenon characterizes wavefunctions at critical points, such as the Anderson metal-insulator transition \cite{evers2008anderson, mott2004metal, schreiber1991multifractal, kohmoto1987critical}. This unique wavefunction geometry, maintaining rich power-law scaling across diverse length scales \cite{jack2021visualizing, halsey1986fractal, martin2010multifractal}, suggests that its intrinsic multi-scale nature is a promising, inherent mechanism for generating effective all-to-all interactions.

%\tcg{In this letter, we demonstrate that the intrinsic geometry of multifractal wavefunctions naturally induces an effective all-to-all interaction. We establish this connection by calculating the multifractal spectrum for three distinct models—the Fibonacci chain, Sierpi\`nki gasket, and the 3D Anderson model—and showing that the resulting interactions are long-ranged and global. This analysis reveals that the multi-scaled nature of these wavefunctions fundamentally overcomes the typical locality constraint of single-scale systems. Furthermore, we show that the quench dynamics of this multifractality-driven all-to-all system exhibit rapid information scrambling, distinct from conventional short-range models, thereby establishing it as an ideal platform for simulating quantum chaos. In particular, the Quantum Fisher Information exhibits an ultrafast decay accompanied by negative values in the Tripartite Mutual Information—a hallmark of fully delocalized entanglement—with the saturation of the upper bound of entanglement velocity. Consequently, our work validates a novel paradigm where the inherent geometric complexity of quantum wavefunction facilitates to investigate global physical phenomena.}

In this Letter, we demonstrate that the intrinsic geometry of multifractal wavefunctions naturally induces an effective all-to-all interaction. We establish this by analyzing the multifractal spectrum \cite{jack2021visualizing, halsey1986fractal, martin2010multifractal} and emergent interactions across three distinct models: the Fibonacci chain, Sierpi\`nki gasket, and the 3D Anderson model \cite{
PhysRevB.111.L020411, jagannathan2021fibonacci, domany1983solutions, wang1996magnetic, vsuntajs2021spectral, shklovskii1993statistics}. Our analysis confirms that the multi-scaled nature of these wavefunctions fundamentally overcomes the locality constraint. Furthermore, we demonstrate that this multifractality-driven system facilitates rapid information scrambling. 
This chaotic behavior is confirmed by the ultrafast shift of the quantum Fisher information, the bipartite mutual information and negative tripartite mutual information \cite{iyoda2018scrambling, seshadri2018tripartite,hosur2016chaos}, proving the onset of highly delocalized entanglement.
Our work validates a novel paradigm where the inherent geometric multifractality of a quantum wavefunction facilitates the investigation of global, highly chaotic physical phenomena.

\begin{figure*}[htb!]
    \centering
    \includegraphics[width=1\linewidth]{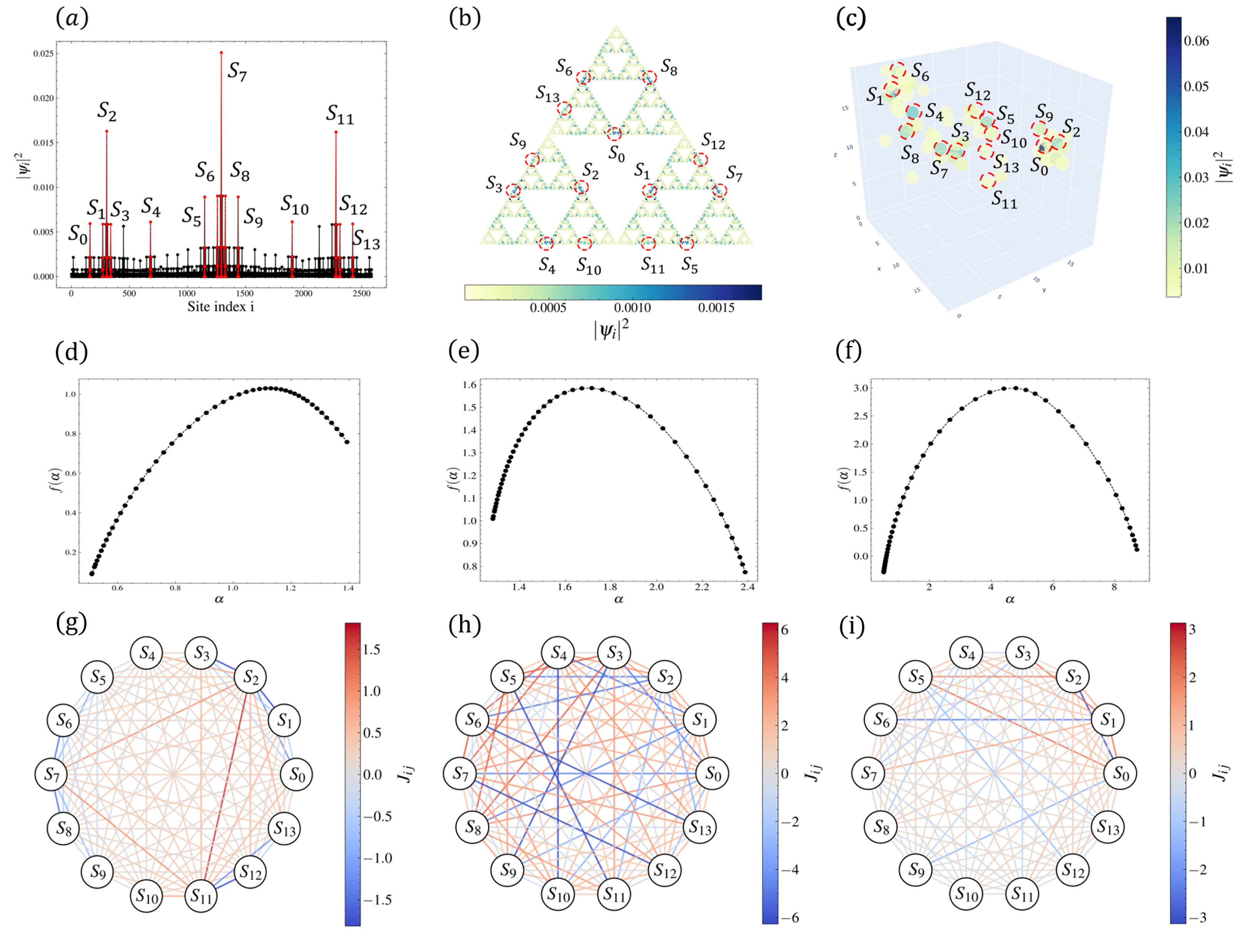}
    \caption{All-to-all interactions and multifractal wavefunctions.
(a,d,g) Fibonacci chain with $t_{A}/t_{B}=0.6$ and $N=2585$;
(b,e,h) Sierpi\`nski gasket of the $7$th generation with magnetic flux $\phi=0.01$;
(c,f,i) three-dimensional Anderson model on a $20\times20\times20$ lattice with disorder strength $W=16.5$.
The hopping amplitude is set to $t=1$ (with $t_{A}=1$ for the Fibonacci chain).
Panels (a-c) show representative single-particle wavefunctions at half filling, where red-highlighted regions labeled $S_i$ denote the $i$th collective spin block.
Panels (d-f) display the corresponding multifractal spectra $f(\alpha)$.
Panels (g-i) present the zero-temperature effective all-to-all interaction strengths $J_{ij}$ between the spin blocks.
}
    \label{fig:all}
\end{figure*}

\textit{\tcm{Multifractal Wavefunctions---}}
To validate the general mechanism linking multifractality to effective all-to-all interactions, we analyze three representative systems that host multifractal wavefunctions. These models span a spectrum of complexity, moving from quasiperiodic order to pure disorder. 
First, we examine the off diagonal Fibonacci chain, described by the Hamiltonian $H_{F}=-\sum_n t_n (c^\dagger_{n+1} c_n+h.c.)$ where $c_{i}^{(\dagger)}$ annihilates (creates) the electron on $i$ site. Here, the hopping amplitude $t_n$ takes one of two values, $t_A$ or $t_B$ according to the $n$th element of Fibonacci sequence, generated by the substitution rule $A\rightarrow AB$ and $B\rightarrow A$. This pattern is known to be characterized by the hyperspace geometry, imposing the additional length scale onto the wavefunction \cite{jagannathan2021fibonacci, PhysRevB.111.L020411}.
Second, we examine the Sierpi\`nski gasket, of which the lattice itself is geometrically fractal. The lattice is constructed by recursive dividing of equilateral triangles. The tight binding Hamiltonian is given by $H_{\mathrm{S}}=t\sum_{\langle i, j\rangle} (e^{i\phi_{ij}}c_i^\dagger c_j+h.c.)$ where $\langle i, j\rangle$ denotes the nearest-neighboring sites \cite{wang1996magnetic, domany1983solutions}. The phase $\phi_{ij}$ is the magnetic flux corresponding to the Peierls substitution, introduced to perturbatively lift the high degeneracy of energy levels. The phase factor $\phi_{ij}$ is selected to have $+\phi(-\phi)$ when the $i, j$ pairing follows the clockwise(counterclockwise) rotation on the up triangle. We set $\phi=0.01$.
Although two previous models have inherent fractality, they both possess the underlying patterns and long-range orders. To finally generalize the argument to system without any orderings, we consider the 3D Anderson model. The Hamiltonian is defined on a cubic lattice as $H_A=-t\sum_{<i, j>}(c_i^\dagger c_j+h.c.) +\frac{W}{2}\sum_i \epsilon_i c^\dagger_i c_i$.
Here each on-site potential $\epsilon_i$ are independent and identical random variables of uniform probability in $[-1, 1]$ and $W$ is the disorder strength. At the critical value $W=W_c \approx 16.5$ the system undergoes the metal-insulator transition and the eigenstates become multifractal \cite{vsuntajs2021spectral,shklovskii1993statistics}. Focusing on the zero-temperature limit, our approach utilizes a single, fixed realization of quenched disorder, without performing a disorder average.

To quantify the multifractality of a wavefunction, we introduce the multifractal spectrum $f(\alpha)$. Let us consider the boxes of size $L$, partitioning the given arbitrary probability distribution with newly defined box probability $P_i(L) = \int_{\text{box } i} |\psi(\mathbf{r})|^2 d\mathbf{r}$. 
From this probability distribution, we define the generalized inverse participation ratio (IPR) with $q$-th moment as $\mathrm{IPR}_q = \sum_i P_i(L)^q$. The scaling of this quantity with system size is characterized by exponents $\tau(q)$, defined by the relation $\mathrm{IPR}_q \sim L^{\tau(q)}$ in the limit $L \rightarrow 0$.
The full multifractal spectrum, $f(\alpha)$, is obtained via a Legendre transform on the function $\tau(q)$ with respect to $q$: $f(\alpha) = q\alpha - \tau(q)$
where $\alpha = d\tau(q)/dq$ \cite{jack2021visualizing, halsey1986fractal, martin2010multifractal}. It describes the average scaling exponent of the local probability ($P_i(L) \sim L^\alpha$), while $f(\alpha)$ represents the fractal dimension characterized by that exponent. A non-trivial spectrum for $f(\alpha)$ is the hallmark of a multifractality. The width of this spectrum serves as a direct measure of the wavefunction's complexity and the richness of its spatial scalings. In contrast, monofractal states such as fully localized or extended states are described by a single point in the $f(\alpha)$ spectrum ($\alpha=0, f(0)=0$ and $\alpha=1, f(1)=1$ respectively in 1D). 

Our analysis focuses on the wavefunctions at half-filling, which we identify as the effective Fermi level for convenience. Figures \ref{fig:all}(a–c) show representative eigenstates from the three systems. The Fibonacci chain and the Sierpiński gasket display characteristic spatial patterns reflecting the long-range order of their underlying quasiperiodic or fractal geometry, whereas the eigenstates of the 3D Anderson model appear irregular and structureless due to the presence of strong disorder (for clarity, probabilities below 0.005 are omitted in the Anderson model plot). 
Figures \ref{fig:all} (d-f) show the corresponding multifractal spectra $f(\alpha)$. All three models possess a rich distribution of scaling exponents $\alpha$, and the maximum value of $f(\alpha)$ occurs at a non-integer valued $\alpha$. Both features are hallmark indicators of multifractality. Consistent with the visual structure of the wavefunctions, the 3D Anderson model exhibits the widest $\alpha$ spectrum, reflecting the strong randomness and high dimensionality. The Sierpi\'nski gasket and Fibonacci chain show comparatively narrower, but still clearly multifractal, spectra.

\textit{\tcm{Effective All-to-all Interaction---}}
To probe magnetic responses mediated by multifractal electrons, we introduce a local Kondo-type coupling between the electronic spins $\vec{s}_i$ and localized impurity moments $\vec{S}_{i}$, $H_K = J_K \sum_i \vec{S}_{i}\cdot \vec{s}_i$, where $J_{K}$ is a coupling strength, $\vec{s}_i=c_{i}^\dagger \vec{\sigma}c_i/2$ is the spin of electron and $\vec{\sigma}$ are Pauli matrices.
In the weak-coupling regime $\vert J_K\vert\ll 1$, integrating out the electrons perturbatively yields an effective indirect interaction between impurities, $\mathcal{H}_{\text{eff}}=\sum_{i\neq j} J_{ij}\vec{S}_i\cdot\vec{S}_j$, where $J_{ij}$ encodes the electron-mediated exchange \cite {PhysRev.96.99, PhysRevB.36.3948}. 
To analyze the effective connectivity, we identify neighboring regions of spins exhibiting predominant ferromagnetic coupling and group them into collective spin blocks (see Fig. \ref{fig:all}(a-c)). This mapping effectively approximates the original lattice problem into a network of collective spins. 
For each model, we select 14 ferromagnetic spin blocks according to a criterion of enhanced local wavefunction weight, ensuring that the resulting blocks correspond to physically relevant, strongly correlated regions.
Figures \ref{fig:all}(g-i) illustrate the resulting effective all-to-all interactions between these blocks, obtained by summing all microscopic inter-block couplings $J_{ij}$. It reveals that a connectivity of multi-scale systems fundamentally deviates from the local constraints of single-scale cases. 
Unlike short-range interactions with rapid spatial decay, or long-range RKKY-type couplings that decay in an oscillatory manner, the effective interactions in multifractal systems persist across large spatial separations, resulting in robust global, effectively all-to-all connectivity.
%Furthermore, because this all-to-all interaction emerges between collective spin blocks, it is not confined to fine-tuned individual sites but instead represents a generic and robust feature of the collective regime.
Crucially, this nonlocal connectivity originates directly from the intrinsic geometry of the underlying wavefunctions and persists without the need for external engineering or parameter fine-tuning.

%\tcg{Figures \ref{fig:all} (g-i) illustrate the effective all-to-all interaction between spin blocks. Unlike conventional monofractal systems where spins predominantly interact with their real space neighbors, the spin blocks in multifractal models couple to a large portion of other blocks regardless of distance. It is evident that multiple scaling factors arising from the wavefunction geometry distort the locality constraint and impose global, all-to-all connectivity. Furthermore, the additional factors in self-similarity of the Sierpi\'nski lattice itself results in many equivalently strong all-to-all interactions whereas the others exhibit a more widely distributed range of interaction strengths. Notably, as long as the system resides at the critical point of the metal-insulator transition, multifractality is a pervasive feature of the eigenspectrum. Consequently, the emergence of all-to-all physics is robust and parameter-free, solely depending on the intrinsic geometry of wavefunctions.}

\begin{figure}
    \centering
    \includegraphics[width=1\linewidth]{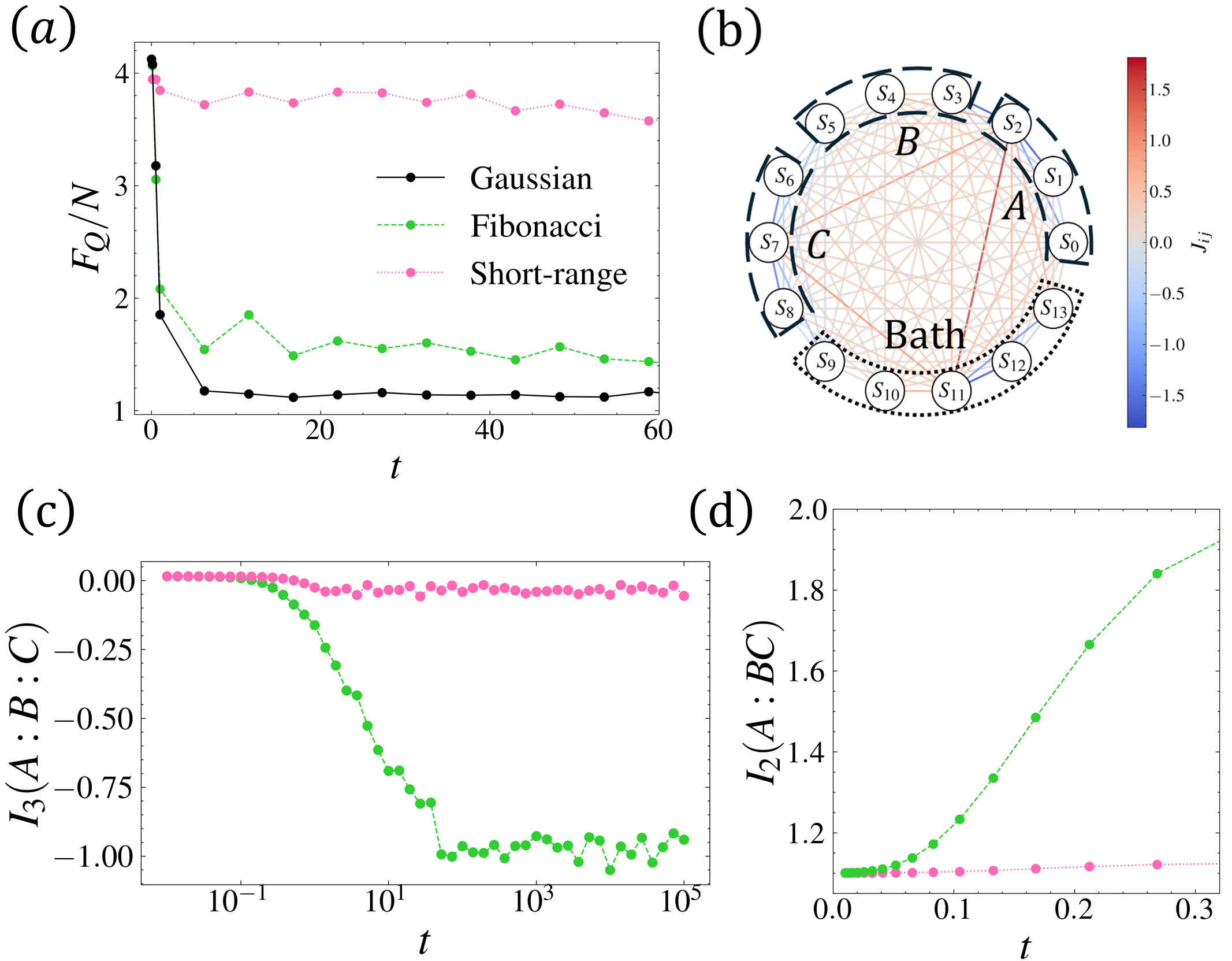}
    \caption{Time evolution of various entanglement diagnostics following a quench to the effective interaction derived from the Fibonacci chain (green), compared with a short-range Heisenberg model with nearest-neighbor coupling $J_{1}$ and next-nearest-neighbor coupling $J_{2}=J_{1}/5$ (pink), and a Gaussian random interaction with zero mean and standard deviation $0.5$, averaged over 10 disorder realizations (black).
(a) Quantum Fisher Information associated with the staggered magnetization operator.
(b) Schematic of the subsystem partitioning for the Fibonacci all-to-all system.
(c) Tripartite mutual information among subsystems $A$, $B$, and $C$.
(d) Bipartite mutual information between subsystem $A$ and subsystem $BC$.} 
    \label{fig:Scrambling}
\end{figure}

\textit{\tcm{Quench Dynamics and Rapid Information Scrambling---}}
One of the key motivations for studying all-to-all interacting spin systems is their ability to reproduce fast information scrambling and quantum-chaotic dynamics reminiscent of those explored in SY-type models, due to non-local entanglement \cite{sekino2008fast, PhysRevB.100.115150, hosur2016chaos, maldacena2016remarks}. 
Here, we demonstrate that this multifractal-induced all-to-all coupling exhibits similarly rapid entanglement dynamics, as evidenced by the ultrafast growth of various entanglement measures following a quench  \cite{bandyopadhyay2023universal}.
%Furthermore, we show that the observed growth rate approaches the theoretical upper bound for entanglement generation in locally unconstrained spin systems, indicating that multifractality enables a scrambling efficiency comparable to that of engineered all-to-all models.

We prepare an initial state $\vert \psi(0) \rangle$ and perform a quench by evolving the system under the effective all-to-all Hamiltonian $\mathcal{H}=\sum_{i\neq j} J_{ij}\vec{S}_i\cdot\vec{S}_j$ as $\vert \psi(t) \rangle=e^{-i\mathcal{H}t} \vert \psi(0) \rangle$. 
It is expected that genuine all-to-all system will exhibit a rapid change in an entanglement measure under such time evolution. Accordingly, we first focus on the Quantum Fisher Information (QFI), experimentally accessible quantity via structure factor \cite{scheie2021witnessing, hauke2016measuring, sabharwal2025witnessing, menon2023multipartite, lovesey1984theory}, as our primary measure for diagnosis. Particularly for a pure state, QFI is defined as, 
\begin{align}
    \label{QFI}
    F_{Q}[\vert \psi(t) \rangle ,\hat{O}]= 4\left(  \langle \psi(t) \vert  \hat{O}^{2}\vert \psi(t) \rangle - \left\vert \langle \psi(t) \vert  \hat{O}\vert \psi(t) \rangle\right\vert^{2}\right)  
\end{align}
where $\hat{O}$ is an observable embedding unitary transformation on the system. Thus, QFI measures how sensitively a quantum state responds to an infinitesimal unitary transformation generated by a given operator. 
We define the staggered magnetization operator as $\hat{O}=\sum_{i}(-1)^{i}\hat{S_{i}^{z}}$ where $\hat{S_{i}^{z}}$ is a $z$ component of $i$th-spin. The system is initialized in the ground state of the one-dimensional antiferromagnetic ($J>0$) Heisenberg chain with periodic boundary conditions, governed by the Hamiltonian
$H=J\sum_{\langle i, j\rangle}\vec{S}_i\cdot\vec{S}_j$ with periodic boundary conditions. The ground state of this model realizes a gapless Luttinger-liquid phase \cite{bethe1931theorie,affleck1988field,imambekov2012one, giamarchi2003quantum}, characterized by algebraically decaying spin correlations and strong multipartite quantum entanglement. As a consequence, the QFI associated with the staggered magnetization operator is large, providing a sensitive witness of many-body entanglement in the ground state \cite{pezze2009entanglement,pezze2018quantum}.

Figure~\ref{fig:Scrambling}(a) shows the time evolution of the quantum Fisher information (QFI).
We consider a one-dimensional Fibonacci chain as a representative example of a multifractal system. For comparison, we analyze a Heisenberg model with Gaussian-distributed all-to-all interactions, as well as a short-range spin chain with only nearest-neighbor ($J_1$) and next-nearest-neighbor ($J_2$) couplings.
We find that the effective all-to-all interaction emergent in the Fibonacci chain leads to a rapid suppression of the QFI at early times, signaling a fast loss of local information about the initial state. Such a rapid decay of QFI is a hallmark of information scrambling, reflecting the delocalization of initially accessible information over many degrees of freedom \cite{bandyopadhyay2023universal}.
At longer times, the QFI in the Fibonacci chain saturates at a comparatively lower value, indicating that while information is efficiently scrambled, the system retains a degree of robustness against complete thermalization or strong perturbations. Notably, this dynamical behavior closely parallels that observed in the Heisenberg chain with Gaussian all-to-all interactions, highlighting the effective global connectivity induced by multifractal wave functions.
By contrast, the QFI in models with predominantly short-range interactions exhibits a slow decay accompanied by pronounced temporal fluctuations over long timescales, indicating strongly suppressed information scrambling.

To further validate the emergence of nonlocal entanglement, we introduce the tripartite mutual information (TMI) under the same quench dynamics used to analyze the QFI. The TMI for three subsystems $A$, $B$ and $C$ is defined as, \cite{iyoda2018scrambling, seshadri2018tripartite,hosur2016chaos}
\begin{align}
    \label{TMI}
    I_{3}(A:B:C) = I_{2}(A:B) + I_{2}(A:C) - I_{2}(A: BC). 
\end{align}
$I_{2}(A:B) = S_{A}+S_{B}-S_{AB}$ indicates the Bipartite Mutual Information (BMI), where $S_{A}=-\mathrm{Tr}(\rho_{A}\ln \rho_{A})$ is the von Neumann entropy of the reduced density matrix $\rho_{A}=\mathrm{Tr}_{A^{c}}(\rho)$ with  $A^{c}$ the complement of subsystem $A$. The BMI quantifies the total (classical and quantum) correlations shared between subsystems  $A$ and $B$.
A negative value of the TMI provides a strong diagnostic of information scrambling \cite{iyoda2018scrambling}. Physically, it signifies that information initially localized in subsystem 
$A$ is not stored in either $B$ or $C$ individually, but is instead delocalized into genuinely multipartite correlations spanning the composite system. In this regime, recovery of information about $A$ requires joint access to the combined subsystem $BC$, rather than to $B$ or $C$ alone.

%Figure \ref{fig:Scrambling}(c) presents the TMI dynamics for the Fibonacci chain, partitioned into three subsystems $A$, $B$, $C$, and a bath (see Fig.\ref{fig:Scrambling}(b)). The TMI of the multifractal system saturates to a significantly lower negative value compared to the short-range model, confirming that information is genuinely scrambled into non-local degrees of freedom rather than beging retained in local correlations. Figure \ref{fig:Scrambling}(d) displays the dynamics of the BMI between subsystem $A$ and the composite system $BC$. We observe that this BMI increases rapidly, signaling the rapid construction of multipartite entanglement across all three regions. This rapid delocalization explains the origin of the observed QFI decay and the negative TMI.
%Collectively, these measures confirm that multifractality-driven all-to-all systems function as ultrafast scramblers, demonstrating their potent utility for simulating chaotic quantum dynamics.

Figure~\ref{fig:Scrambling}(c) shows the time evolution of the tripartite mutual information (TMI) for the Fibonacci chain, where the system is partitioned into three subsystems 
$A$, $B$, $C$, together with a bath (see Fig.~\ref{fig:Scrambling}(b)). We find that the TMI of the multifractal system saturates to a significantly more negative value than that of the short-range model, indicating that information is genuinely scrambled into nonlocal degrees of freedom rather than being retained within local or pairwise correlations.
Figure~\ref{fig:Scrambling}(d) presents the dynamics of the bipartite mutual information (BMI) between subsystem 
$A$ and the composite subsystem $BC$. The BMI grows rapidly at early times, signaling the fast buildup of multipartite entanglement shared across all three subsystems. This rapid delocalization of information provides a microscopic explanation for both the observed suppression of the QFI and the emergence of negative TMI.
Taken together, these diagnostics demonstrate that multifractality-induced effective all-to-all interactions lead to ultrafast information scrambling, highlighting the utility of such systems as controlled platforms for exploring strongly nonlocal and chaotic-like quantum dynamics.

\textit{\tcm{Conclusion---}}In summary, we have identified a novel mechanism by which the intrinsic geometry of multifractal wavefunctions mediates effective all-to-all spin–spin interactions. By systematically analyzing three representative systems, the Fibonacci chain, the  Sierpi\'nski gasket, and the three-dimensional Anderson model, we demonstrate that multifractality generically enables indirect exchange processes with global connectivity, thereby circumventing the locality constraints that typically govern conventional condensed-matter systems.
The genuinely all-to-all character of these interactions is further substantiated through quench dynamics of entanglement diagnostics. In particular, the rapid decay of the quantum Fisher information, the saturation of the tripartite mutual information to negative values, and the ultrafast growth of the bipartite mutual information collectively establish these systems as fast scramblers of quantum information.

Our results position multifractal wavefunctions as a generic and controllable platform for exploring many-body quantum chaos and nonlocal collective phenomena. Looking forward, extending this framework to experimentally tunable settings, such as engineering multifractal electronic structures via twist-angle control in multilayer van der Waals materials \cite{andrei2020graphene,lopes2007graphene,bistritzer2011moire,ren2025moire}, or exploiting recently synthesized van der Waals quasicrystals \cite{tokumoto2024superconductivity,terashima2024anomalous}, offers a promising route toward realizing tailored multifractality and simulating exotic non-equilibrium dynamics in experimentally accessible quantum matter\footnote{Y. Joung, J. Park, and S. B. Lee, manuscript in preparation.}.

\textit{\tcm{Acknowledgement---}}This work was supported by the National Research Foundation of Korea (NRF) under Grant No. 2021R1A2C1093060, and by the Nano Material Technology Development Program through the National Research Foundation of Korea (NRF), funded by the Ministry of Science and ICT, under Grant No. RS-2023-00281839.

\bibliography{reference.bib}
\end{document}